\documentclass[article,shortnames]{jss}
\usepackage{color}
\usepackage{booktabs}
\usepackage[singlelinecheck=false,
            justification=RaggedRight,
            format=default,
            labelfont=bf,
            font={small}
           ]{caption}[2005/06/28]
\usepackage{acronym}
\usepackage{enumitem}
\usepackage{xspace}

\author{Lukas W.~Lehnert \And 
        Hanna Meyer \And
        Wolfgang A.~Obermeier \AND
        Brenner Silva \And
        Bianca Regeling\And 
        Boris Thies \AND
        J\"{o}rg Bendix\vspace{.2cm}\\Faculty of Geography\\Philipps-University of Marburg
}
\title{Hyperspectral Data Analysis in R: The \pkg{hsdar}-Package}

\Plainauthor{Lukas W.~Lehnert, Hanna Meyer, Wolfgang A.~Obermeier, Brenner Silva, Bianca Regeling, J\"{o}rg Bendix} 
\Plaintitle{Hyperspectral Data Analysis in R: the hsdar Package} 
\Shorttitle{Hyperspectral data analysis in R} 

\Abstract{
  Hyperspectral remote sensing is a promising tool for a variety of applications including ecology, geology, analytical chemistry and medical research. This article presents the new \hsdar package for \proglang{R} statistical software, which performs a variety of analysis steps taken during a typical hyperspectral remote sensing approach. The package introduces a new class for efficiently storing  large hyperspectral datasets such as hyperspectral cubes within \proglang{R}. The package includes several important hyperspectral analysis tools such as continuum removal, normalized ratio indices and integrates two widely used radiation transfer models. In addition, the package provides methods to directly use the functionality of the \pkg{caret} package for machine learning tasks. Two case studies demonstrate the package's range of functionality: First, plant leaf chlorophyll content is estimated and second, cancer in the human larynx is detected from hyperspectral data.
}
\Plainkeywords{hyperspectral remote sensing, hyperspectral imaging, spectroscopy, continuum removal, normalized ratio indices}

\Address{
  Lukas W.~Lehnert\\
  Laboratory for Climatology and Remote Sensing\\
  Faculty of Geography\\
  Philipps-University of Marburg\\
  35037 Marburg, Germany\\
  E-mail: \email{lukas.lehnert@staff.uni-marburg.de}\\
  URL: \url{http://www.lcrs.de}
}

\newcommand{\hsdar}{\pkg{hsdar}\xspace}

\setkeys{Gin}{width=.75\linewidth}

\begin{document}

\newacro{RMSE}[RMSE]{root mean square error}

\section{Introduction}

Hyperspectral data refers to measurements of reflectance, transmission or absorption of electromagnetic radiation with a very high spectral resolution. Consider photographs taken with a normal digital camera to illustrate the concept of spectral resolution. The sensors in digital cameras have three bands that cover the blue, green and red portions of the visible electromagnetic radiation. Each band is sensitive to radiation in a wavelength range of approximately 100~nm. 
Hyperspectral sensors, in contrast, feature hundreds of such bands that are sensitive to a very narrow wavelength range along the electromagnetic spectrum (often down to 1~nm). Together, all bands continuously cover a certain portion of the electromagnetic spectrum. Additionally, most hyperspectral sensors feature bands within the infrared or ultraviolet ranges. For instance, the hyperspectral satellite sensor Hyperion provides data with 220 bands with a spectral resolution of approximately 11~nm (wavelength range) at each 10~nm (sampling interval) from 400~nm (visible) to 2500~nm (short-wavelength infrared, \citealp{Pearlman}). 

Hyperspectral imaging, also referred to as imaging spectroscopy, is used in various disciplines, such as analytical chemistry \citep{Blanco2002}, agricultural research (precision farming, \citealp{Haboudane2002}), ecology \citep{Ustin2004}, pedology \citep{Gomez2008}, geology \citep{Bishop2011}, and medical research \citep{Calin2014,Regeling2015}. The main advantages of hyperspectral imaging are its cost-effectiveness in spatial analysis, the non-destructive measurement of biophysical and biochemical properties of the investigated surface and the speed of analysis (up to real-time). Hyperspectral analysis is not restricted to space-born approaches. Many of the above-mentioned fields make use of portable spectrometers or hyperspectral cameras, which can be used in the field, in the laboratory or even in a surgical suite. The choice of the measuring device and its spectral specifications depends on the surface under investigation and the aim of the analysis. For instance, vegetation has a very prominent spectral feature called the red-edge. This refers to a sharp increase of reflectance values in the near infrared wavelengths. These wavelengths, in contrast, are less informative in geological analyzes, which usually require the short- and mid-infrared wavelengths. 

Currently, most hyperspectral approaches use commercial software tools such as Erdas Imagine, \proglang{ENVI} or the hyperspectral toolbox in \proglang{MATLAB}. These tools are generally expensive and have limited functionalities for statistical analysis. Therefore, we developed a new package in the open source software \proglang{R} \citep{RDevelopmentCoreTeam2010}. The \textbf{H}yper\textbf{s}pectral \textbf{D}ata \textbf{A}nalysis (``\pkg{hsdar}'') package combines important hyperspectral analysis tools with the statistical power of \proglang{R}. This article is structured as follows: The first section summarizes the reasons why \proglang{R} is convenient for hyperspectral analysis. The next section outlines  the main functionalities and the implementation of the \pkg{hsdar} package, and also compares it with other available software tools with a special focus on the other hyperspectral package \pkg{``hyperSpec''} in \proglang{R}. Finally, two examples demonstrate the effectiveness of combining hyperspectral techniques with the statistical power of \proglang{R}.

\section[Why use R for hyperspectral imaging analysis]{Why use \proglang{R} for hyperspectral imaging analysis}
The methodology which is commonly applied in the analysis of hyperspectral datasets consists of three parts: (1) the preprocessing of spectra, (2) the extraction of the relevant information (i.e.,~spectral characteristics associated with biophysical properties of the target), and (3) a classification or regression analysis to predict biophysical properties in space and time. \proglang{R} is the most comprehensive software tool for performing statistical analyses during step (3). In this context, especially the machine learning algorithms such as support vector machines, Random forests and artificial neural networks are powerful tools for modelling different parameters across space and time (for applications see e.g., \citealp{Schwieder2014,Hansen2002,Bacour2006}). However, the functionality required for steps (1) and (2) has only been partly available in \proglang{R}, was distributed across multiple packages and was not directly applicable to hyperspectral data. 

Thus, to take advantage of the statistical power of \proglang{R} for hyperspectral data analysis, a new package was developed that provides a framework for handling and analyzing hyperspectral data. A special focus was set on the analysis of large datasets taken under field conditions for e.g., vegetation remote sensing. The \proglang{R}-package \hsdar implements commonly used processing routines for hyperspectral data and further combines or extends the existing functionality of \proglang{R} to include hyperspectral data into a broad range of statistical analyses.

\section[Overview of the functionality of hsdar]{Overview of the functionality of \hsdar}
This section gives a brief technical overview on the general functionality provided by \pkg{hsdar}. The description starts with a short introduction of the classes followed by a summary of the main functions.
\subsection{Classes}
\begin{figure}
\centering
\includegraphics[width=1\textwidth]{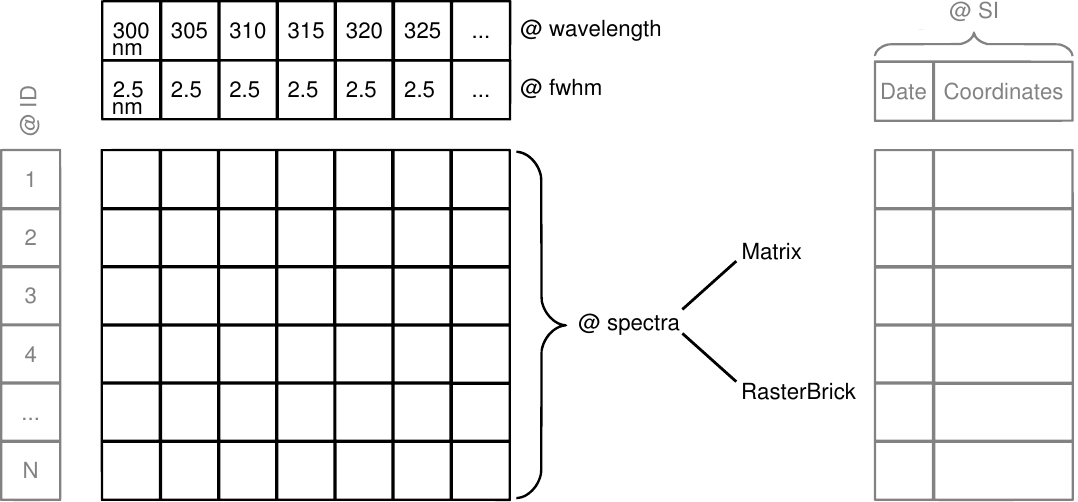}
\caption{Scheme of the S4-class ``Speclib'' Implemented in \hsdar. Black Slots are Required and Grey Ones are Optional. The Spectra's Slot can Either be a Matrix or a RasterBrick Object. The SI Slot can Encompass Various Types of Objects Including Raster Images. Note that Functions Exist to set and return Data in Each Slot.\label{fig:Speclib}}
\end{figure}

To provide a framework to handle large hyperspectral datasets, the \hsdar-package defines a new S4-class called ``Speclib''. This allows the user to store hyperspectral measurements and all information associated with those measurements in a single object (Figure \ref{fig:Speclib}). The hyperspectral measurements consist of reflectance values stored in the spectra slot and their spectral specifications. The spectra are stored either as a numeric matrix or a RasterBrick-object. The matrix is intended for smaller data sets such as point measurements, whereas the RasterBrick object may contain large hyperspectral (satellite) images. If the spectra are stored as a matrix, the rows delineate between different samples while the columns represent the different spectral bands. The spectral specification consists of two numeric vectors stored in the wavelength and the \textbf{f}ull-\textbf{w}idth-\textbf{h}alf-\textbf{m}aximum (fwhm) slots.  The wavelength gives the central position of each band and the fwhm value describes the difference between the wavelength values where the sensitivity of the sensor is half of its maximum in the respective band. Both values are specifications of the sensor used to acquire the data and must be in the same unit. It is preferred to use $\textrm{nm}$ but automatic conversion from other typical units such as $\mu\textrm{m}$ is supported. If the fwhm values are unknown, the difference between neighboring bands are used as an approximation. The associated data (termed SI as an abbreviation for \textbf{s}upplementary \textbf{i}nformation), which is included as a list, may contain any type of ancillary information like the measurement setup or the geographical position. Additionally, raster images are supported as part of the SI.  

Speclibs can be created through several methods. For each method, the user must at least know the wavelength values of all bands that must be available as a numeric vector. The most important method to create an object of class Speclib is using the file path pointing to a hyperspectral raster image  readable by \pkg{rgdal} or \pkg{raster} \citep{Hijmans2016,Bivand2016,Pebesma2015}. The second option to create a Speclib is to read the reflectance values from a file (e.g., a comma-separated list) and store these in a matrix. This matrix, together with the wavelength information, can then be used to create a Speclib. In the following short example, the example dataset ``spectral\_data'' (which is already a Speclib) is divided into its basic components, which are then used to create a new Speclib:
\begin{Schunk}
\begin{Sinput}
R> library("hsdar")
R> data("spectral_data")
R> reflectance <- spectra(spectral_data)
R> class(reflectance)
\end{Sinput}
\begin{Soutput}
[1] "matrix"
\end{Soutput}
\begin{Sinput}
R> wv <- wavelength(spectral_data)
R> class(wv)
\end{Sinput}
\begin{Soutput}
[1] "numeric"
\end{Soutput}
\begin{Sinput}
R> spec_lib <- speclib(reflectance, wv)
R> class(spec_lib)
\end{Sinput}
\begin{Soutput}
[1] "Speclib"
attr(,"package")
[1] "hsdar"
\end{Soutput}
\end{Schunk}
In this example, the spectra (\code{reflectance}) are stored as a matrix and the wavelength (\code{wv}) is stored as a numeric vector.

Aside from using local offline data, \hsdar can search online hyperspectral databases and automatically download data. The following example searches for spectra from grass species in the  \href{https://speclab.cr.usgs.gov/spectral.lib04/spectral-lib04.html}{USGS Digital splib04 Spectral Library} and downloads the data. Note that missing data in the downloaded spectra are automatically masked out.
\begin{Schunk}
\begin{Sinput}
R> avl <- USGS_get_available_files()
R> grass_spectra <- USGS_retrieve_files(avl = avl, 
+    pattern = "grass-fescue")
\end{Sinput}
\end{Schunk}
In the example above, the first command returns all available spectra. Users can specify a subset of spectra in a search string within the retrieve function (in this case ``grass-fescue''), which is downloaded and converted to a Speclib. Note that the function supports approximate string matching so that entries similar to the search string are found.

\subsection{Functionality}
Along with the new Speclib class, \hsdar includes several methods to summarize, plot, query and replace data in Speclib objects. Since many hyperspectral datasets are available as raster datasets (e.g.,~if acquired by satellite), \hsdar provides a simple interface to the \pkg{raster} package that allows users to read and save data from and to all common raster formats via the \pkg{rgdal} interface \citep{Hijmans2016,Bivand2016,Pebesma2015}. On commonly used hardware, hyperspectral raster datasets often exceed the capacity of the RAM. To overcome this issue, \hsdar provides two processing options for such large datasets. The simpler, less computational effective option is to store the spectra as a RasterBrick object in a Speclib. In this case, the spectra are read into memory only upon request and most of the functions process the spectral data block-wise. In this context, the functions automatically detect if the data should be processed block-wise or if all the data should be read before executing the function. For block-wise computation, the resulting spectra are saved as a temporary raster file and the function returns a new Speclib object pointing to the temporary file. The disadvantage of this option is that if more than one function is applied, the spectra have to be saved and re-read multiple times. Thus, a second option is available, which follows the framework of the \pkg{raster} package but requires the user to be familiar with simple programming tasks in \proglang{R}. Like the \pkg{raster} package, \hsdar provides  \code{writeStart}, \code{getValuesBlock}, \code{writeValues} and \code{writeStop} methods for the Speclib class so that the user can easily process a large dataset by iteratively reading parts (chunks) of the images, passing it through multiple functions and writing the result to a new raster file. Only one reading and writing process is required in this case, which  considerably expedites the analysis. A typical code block would look like the following. To execute it, note that \code{wavelength} needs to be defined and \code{infile} must point to an existing file readable by the \pkg{raster} package. The result will be a new file in the GeoTIFF-format defined by \code{outfile} featuring the same number of bands as the existing file (option 'nl'):
\begin{Schunk}
\begin{Sinput}
R> ra <- speclib(infile, wavelength)
R> tr <- blockSize(ra)
R> res <- writeStart(ra, outfile, nl = nbands(ra), format = "GTiff")
R> for (i in 1:tr$n) 
+  {
+    v1 <- getValuesBlock(ra, row=tr$row[i], nrows=tr$nrows[i])
+    v2 <- ANY_FUNCTION(v1)
+    res <- writeValues(res, v2, tr$row[i])
+  }
R> res <- writeStop(res)
\end{Sinput}
\end{Schunk}
In the loop, function(s) provided by the \hsdar package can be applied to the Speclib \code{v1}. Examples of functions will be discussed in detail in the following sections. The result of the function(s) (termed \code{v2} in this example) is then written to the initially defined file (\code{res}). Note that objects \code{res} and \code{v1} are of class Speclib, while \code{v2} may be a vector, matrix or a Speclib depending on the return value of the functions applied in between. Please read the help files and the corresponding vignette available in the \pkg{raster} package for further information.

The functionality provided by the \hsdar package can be divided into preprocessing, analysis and modelling stages (Table \ref{tab:func}). In the following, we briefly outline the most important features except those that are part of the analysis in the section of case studies.
\begin{table}
\begin{small}
\begin{tabular*}{1\textwidth}{@{}c@{\extracolsep\fill}c@{\extracolsep\fill}c@{}}
\toprule
Preprocessing & Analysis & Modeling\\
\midrule
\begin{minipage}[t]{0.32\textwidth}
\begin{itemize}[leftmargin=12pt]
 \item Filtering
 \item Derivations
 \item Spectral resampling
 \item \textit{Continuum removal} 
\end{itemize}
\end{minipage}
&
\begin{minipage}[t]{0.32\textwidth}
\begin{itemize}[leftmargin=12pt]
 \item Red edge parameters
 \item $\sim$ 100 Vegetation indices
 \item Soil indices
 \item \textit{Normalized ratio indices}
 \item Spectral unmixing
 \item \textit{Feature selection algorithms}
 \item \textit{Extraction of absorption features}
\end{itemize}
\end{minipage}
&
\begin{minipage}[t]{0.32\textwidth}
\begin{itemize}[leftmargin=12pt]
 \item Implementation of the leaf reflectance model PROSPECT and the canopy reflectance model PROSAIL
 \item \textit{Link to machine learning functionality of caret \citep{Kuhn2008}}
\end{itemize}
\end{minipage}\\
\bottomrule
\end{tabular*}
\end{small}
\caption{Summary of the Main Functionalities of the \hsdar-package. Items in Italic are Presented in Detail in the Case Studies Section.\label{tab:func}}
\end{table}

\begin{figure}[!t]
\centering
\resizebox{1\textwidth}{!}{
\includegraphics{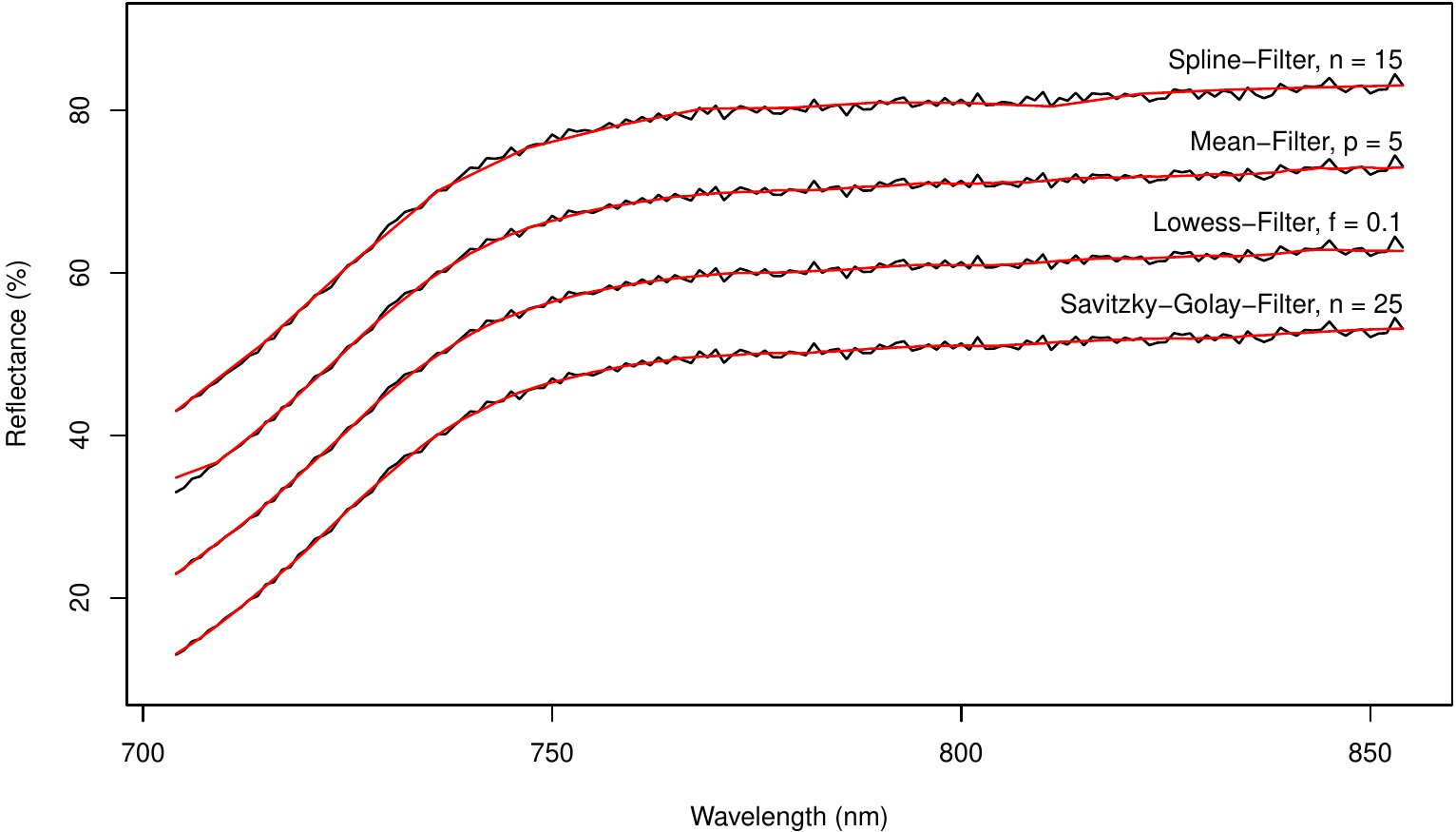}
}
\caption{Effect of Filtering to Reduce Noise in Spectral Data. Red Lines are the Filtered Reflectance and the Black Lines the Raw Reflectance. All Filters are Applied to the Same Spectrum. Note that for Illustration Purposes, the Values of the Lowess-, Mean-, and Spline-Reflectances have been Increased by 10, 20 and 30 \% After Filtering, respectively. Settings for the Filters are as follows: \code{n} and \code{p} for the Savitzky-Golay- Spline- and Meanfilters are the Filter Lengths, whereas \code{f} Gives the Proportion of Bands in the Spectrum that Influence the Smooth at Each Value in the Loewess-filter.\label{fig:filtering}}
\end{figure}

Noise reduction is a critical preprocessing task in hyperspectral analysis because, as a consequence of their high spectral resolution, the sensors often suffer from low signal to noise ratios, thus, an important step of each hyperspectral analysis is filtering the spectra. In \hsdar the function \code{noiseFiltering} applies one of four predefined filters (Savitzky-Golay-, Lowess-, mean-, Spline-filter) or any other filter function from the \pkg{signal} package \citep{Ligges2013}. Figure \ref{fig:filtering} shows the effect of filtering (red lines) spectra that were artificially affected by random noise (black lines).
Additionally, \hsdar provides functions to calculate variables derived from spectral features and allows the user to integrate (bin or spectrally resample) hyperspectral datasets to sensors featuring a lower spectral resolution. Spectral resampling can be performed using predefined spectral response functions of common satellite sensors or using Gaussian spectral response functions defined by the fwhm values of the sensor with the lower resolution. Alternatively, spectral response values may be stored in a Speclib and passed directly to the resampling function.

To analyze hyperspectral datasets, the computation of approximately 100 vegetation and soil indices is implemented in \hsdar. The indices can be accessed via the functions \code{vegindex} and \code{soilindex} which encompass widely used indices such as the normalized difference vegetation index (NDVI, \citealp{Tucker1979}) in addition to specialized indices such as the cellulose absorption index (CAI), which is a proxy for litter amounts and plant coverage \citep{Nagler2003}. Additionally, users can easily define their own index using a simple syntax. In (hyperspectral) remote sensing of vegetation, the sharp increase in the reflectance values between 680 and 750~nm (red edge) is the most important feature, as the shape of the red edge is determined by the amount of water and chlorophyll in the vegetation. Thus, the red edge is seen as a reliable indicator for plant health in addition to leaf area index, plant coverage, chlorophyll, water and nitrogen content (e.g., \citealp{Filella1994}). Different methods for extracting relevant information in the shape of the red edge are included in \hsdar. These encompass common methods such as deriving the red edge inflection point using a Gaussian fit \citep{Miller1990} or more recent advances such as the red edge position through linear extrapolation \citep{Cho2006a}.  Finally, \hsdar provides functionality to perform linear spectral unmixing (LSU, \citealp{Sohn1997}) e.g., for estimating the fractional vegetation cover.

\hsdar implements two frequently used radiative transfer models to simulate the reflectance values of vegetation. The first one is the leaf reflectance model PROSPECT (vers.~5B and D, \citealp{Jacquemoud1990,Feret2017}). The second one is the canopy reflectance model PROSAIL which enhances the functionality of PROSPECT and includes canopy directional reflectance simulation \citep{Jacquemoud2009}. In addition, the inverted PROSPECT model allows the user to estimate the content of various biochemical parameters in the leaves from hyperspectral data \citep{Jacquemoud1993}.

\section{Other hyperspectral imaging tools}
Comparable functionality can be found in commercial software tools, i.e.,~\proglang{MATLAB} (The MathWorks, Inc., Natick, Massachusetts) and \proglang{ENVI} (Environment for visualizing images, Exelis Visual Information Solutions, Boulder, Colorado). A hyperspectral toolbox is available in \proglang{MATLAB} that provides feature extraction algorithms such as principal component analysis as well as supervised classification algorithms such as a Maximum Likelihood classifier \citep{Arzuaga-Cruz2004}. ENVI has functions for preprocessing hyperspectral images such as continuum removal and feature extraction algorithms such as the spectral angle mapper.

In the open source software \proglang{R}, \hsdar completes its hyperspectral functionality together with another major hyperspectral package called \pkg{hyperSpec} \citep{Beleites2016}. The primary difference between the packages is that \hsdar is intended for analyzing datasets collected under field conditions with satellites or spectrometers with a special focus on vegetation and ecosystem remote sensing \citep{Dechant2017,Grose-Stoltenberg2016,Lehnert2014,Meyer2017}. In contrast, the \pkg{hyperSpec} package provides many useful functions for plotting with a special focus on hyperspectral data acquired under laboratory conditions as in chemistry or medical research \citep{Beleites2011,Beleites2013}. Functions in \hsdar allow it to interface with the \pkg{hyperSpec} package, i.e.,~to convert between Speclib objects and the hyperSpec class. Consequently, \hsdar users also have access to various import and plotting functions provided by the latter package. 

\section{Case studies}\label{sec:exmpl}
\begin{figure}[!t]
\textbf{a}\hspace{0.5\textwidth}\textbf{b}\\
\includegraphics[width=0.49\textwidth]{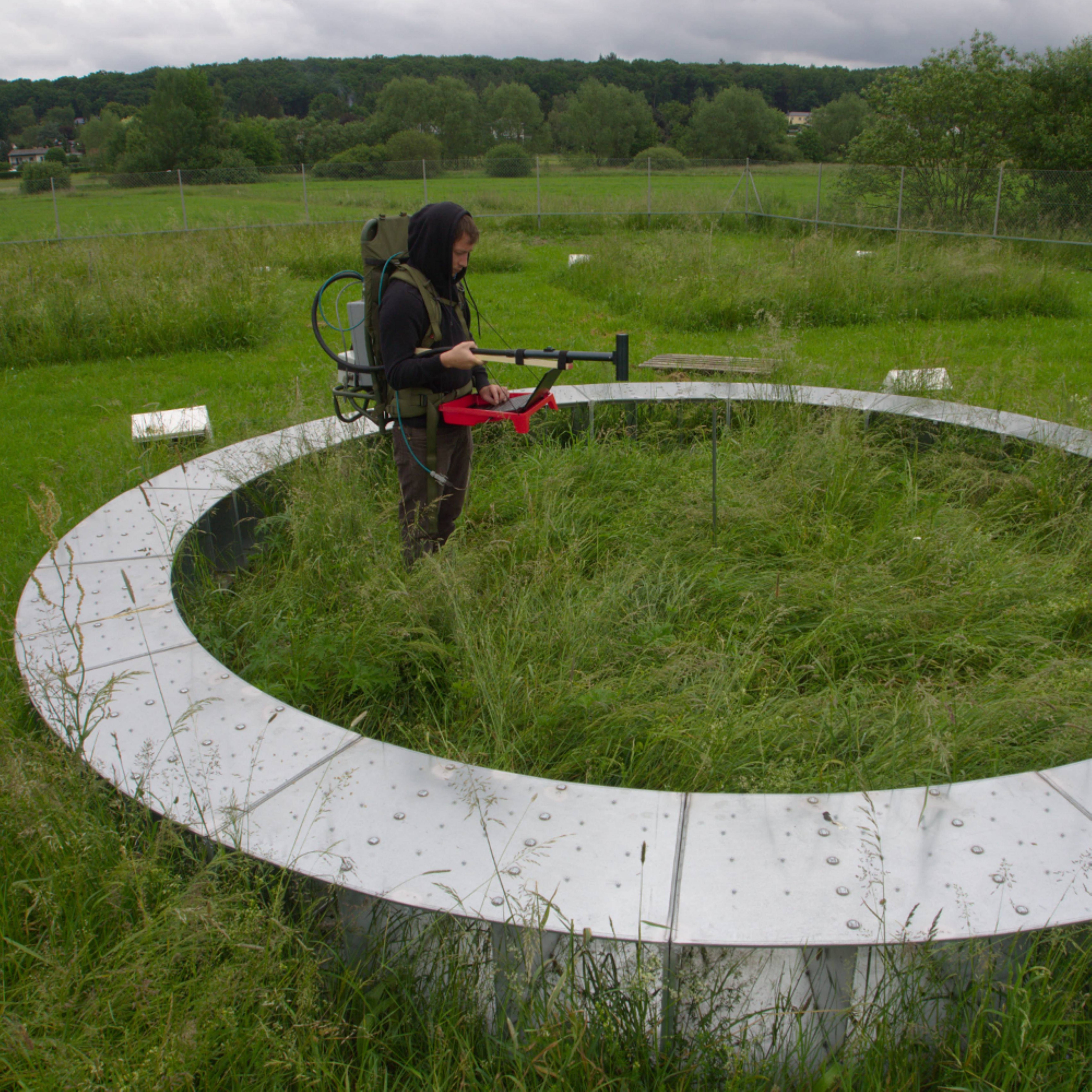}\hspace{0.02\textwidth}\includegraphics[width=0.49\textwidth]{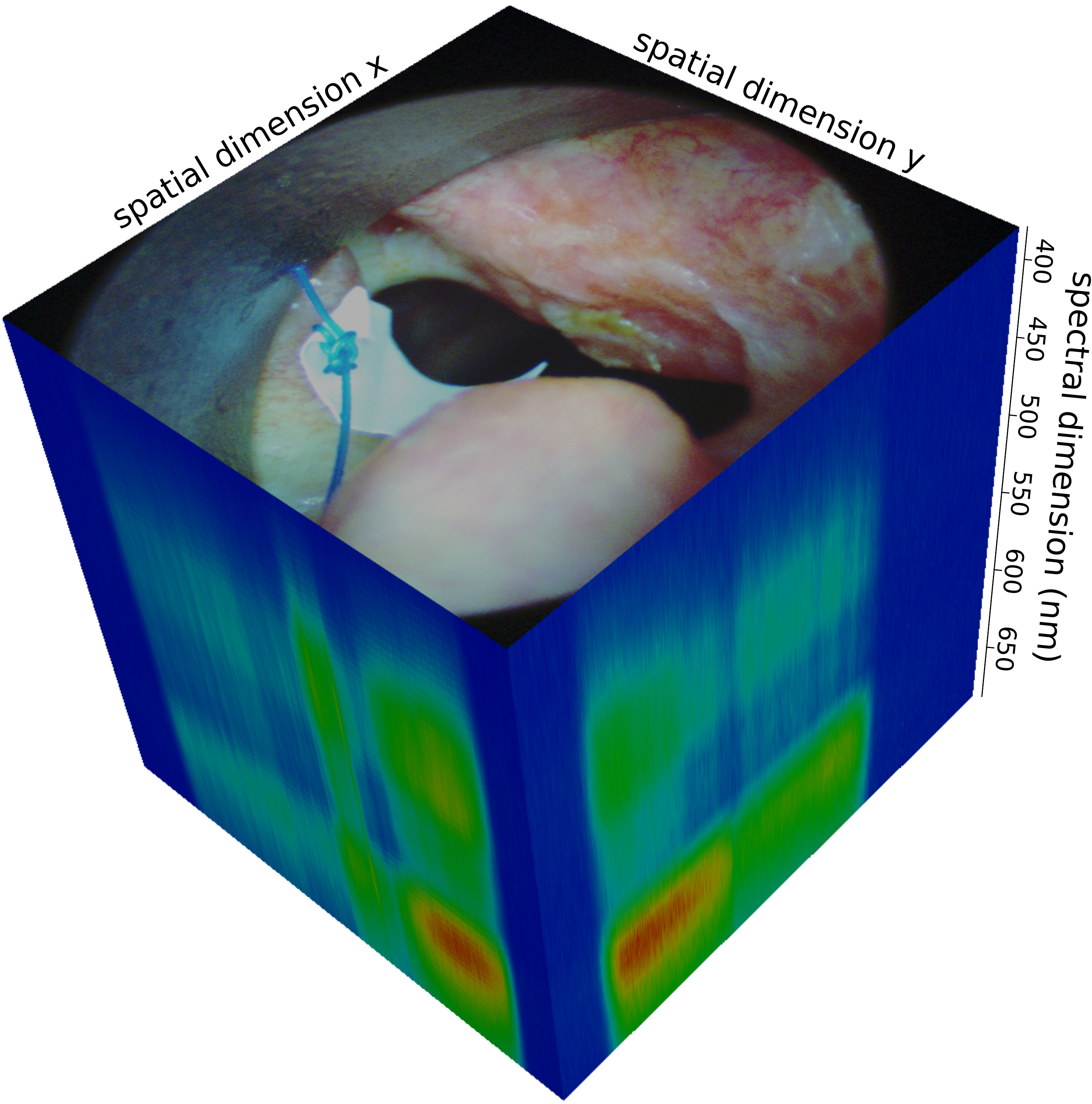}
\caption{Sampling of Hyperspectral Data at the GiFACE Experimental Site with the Spectrometer (a). The Silver Ring is Part of the CO$_2$-Enrichment System. In (b), an Example Image Illustrates the Hyperspectral Cube of the Human Larynx Produced by the \hsdar Function "cubePlot". The RGB-image on top of the Cube is Created from the Bands of the Hyperspectral Image Corresponding to the Center of the Red, Green and Blue Wavelengths. The Colors at the Vertical Sides of the Cube Represent the Intensity Values of the 30 Different Spectral Bands of the Sensor (blue~=~low to red~=~high). \label{fig:imgs_exmpl}}
\end{figure}

In the following sections two study cases are presented to explore the functionality of \hsdar. The first case study uses data from a field experiment conducted in central Germany where hyperspectral images were taken from grassland vegetation exposed to enhanced CO$_2$ air concentrations (Figure \ref{fig:imgs_exmpl}a). The example includes spectra preprocessing, followed by the extraction of absorption features, calibration and validation of a prediction model for chlorophyll content. In the second case study, emphasis is given to the calculation of normalized ratio indices and model parameterization to detect cancer cells in human larynx tissue using hyperspectral images (Figure \ref{fig:imgs_exmpl}b).

\subsection{Remote sensing of vegetation: chlorophyll content}\label{sec:exmpl_cloro}

The first example demonstrates the applicability of \hsdar for hyperspectral data analysis in vegetation studies. Specifically, the package is used to estimate chlorophyll content of plants from hyperspectral data. The dataset was acquired within the scope of a FACE (\textbf{f}ree \textbf{a}ir \textbf{c}arbon dioxide \textbf{e}nrichment) experiment conducted on a temperate grassland situated near Giessen, Germany \citep{Kammann2005,Obermeier2017}. On 15 plots (each 2 x 2 m), the chlorophyll content of the two most abundant grasses (\textit{Arrhenatherum elatius} and \textit{Trisetum flavescens}) was measured using a Konica Minolta SPAD-502Plus chlorophyll meter. The  mean value of chlorophyll content of both species was calculated and weighted by their corresponding plant coverage. Hyperspectral data were acquired at the time of the chlorophyll measurements using a HandySpec\textsuperscript{{\textregistered}} field spectrometer, which simultaneously measures reflectance values from 305~nm to 1705~nm with a spectral resolution of 1~nm (Figure \ref{fig:imgs_exmpl}a). The field spectrometer has two sensors measuring from 305 to 1049~nm and 1050 to 1705~nm. On each plot, 24 spectra were collected under natural (solar) illumination and averaged. Each plot was visited three times, on 30.05.2014, 08.08.2014 and 13.05.2015. Thus, the dataset contains 45 observations. 

The following paragraph describes the preprocessing steps that reduce measurement errors and artifacts in the spectral data. Then, the spectra are transformed to reduce the influence of the illumination at time of acquisition. Finally, the chlorophyll content is estimated with Random Forest using the transformed spectra as predictors \citep{Breiman2001}. Here, we use the \pkg{randomForest} package by \cite{Liaw2002} in combination with the \pkg{caret} package created by \cite{Kuhn2008}.

In the first preprocessing step noise is removed from the spectra using a Savitzky-Golay filter (method ``sgolay'') with a length of 15~nm. The filter reduces the noise of the reflectance values by fitting a polynomial function and eliminates small differences between neighboring bands, which are most likely a result of measurement inaccuracy. 
\begin{Schunk}
\begin{Sinput}
R> data("spectral_data")
R> spectral_data <- noiseFiltering(spectral_data, method = "sgolay", 
+    p = 15)
\end{Sinput}
\end{Schunk}

The result is a Speclib object, which contains a filtered spectral signature in the original sampling  resolution. In addition, the empirical function of \cite{Coste2010} is used to transform the chlorophyll SPAD values to $\mu$g cm$^{-2}$ ($C_{a,b}$) to facilitate the interpretation of the chlorophyll content values:

\begin{equation}
C_{a,b}=\frac{117.1 \cdot\mathit{SPAD}}{148.84 - \mathit{SPAD}}
\end{equation}
Note that the SPAD chlorophyll value is shipped with the example dataset and stored in the supplementary information (SI) of the object.
\begin{Schunk}
\begin{Sinput}
R> SI(spectral_data)$chlorophyll <- 
+    (117.1 * SI(spectral_data)$chlorophyll) /
+    (148.84 - SI(spectral_data)$chlorophyll)
\end{Sinput}
\end{Schunk}

\begin{figure}
\resizebox{1\textwidth}{!}{
\includegraphics{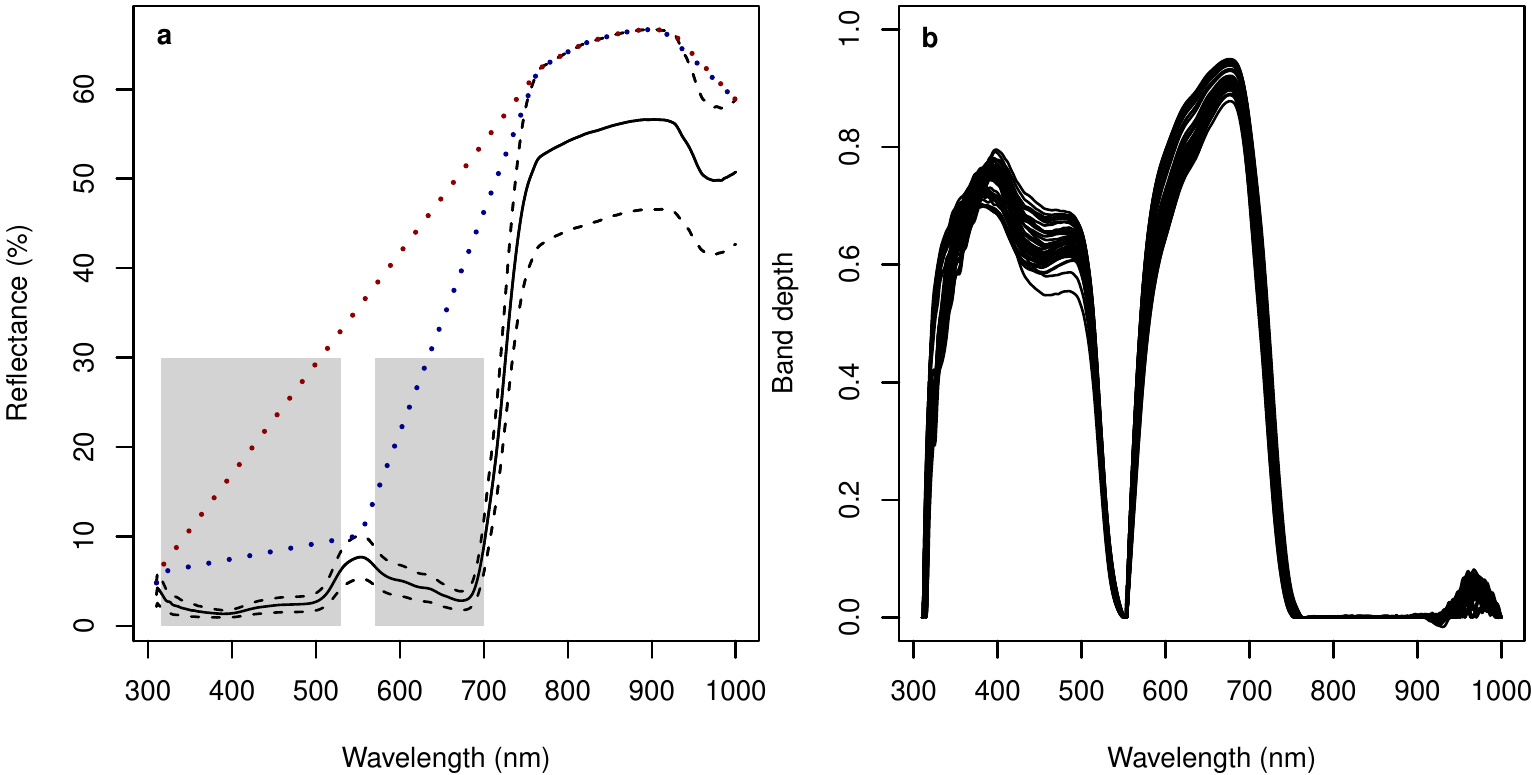}
}
\caption{Spectral Data of the Vegetation at the 15 Plots. Black Lines Show Mean (Solid) and Mean $\pm$ one Standard Deviation (Dashed) of Reflectance Values (a). The Red and Blue Dashed Lines Symbolize the Convex and Segmented Upper Hull of the Upper Standard Deviation Spectrum, respectively. The Gray Boxes Symbolize the Absorption Wavelength of Chlorophyll.  In (b) Band Depth Values are Plotted as the Result of the Segmented Upper Hull Transformation Applied to the Reflectance Spectra. \label{fig:chloro_cr}}
\end{figure}

Chlorophyll strongly absorbs light at around 460~nm in the blue and around 670~nm in the red parts of the electromagnetic radiation (e.g.,~\citealp{Mutanga2004a}). Therefore, the spectra are trimmed to their visible and near infrared part (310 - 1000~nm). The resulting spectral data after preprocessing are visualized in Figure \ref{fig:chloro_cr}a.
\begin{Schunk}
\begin{Sinput}
R> spectral_data <- spectral_data[, wavelength(spectral_data) >= 310 & 
+    wavelength(spectral_data) <= 1000]
\end{Sinput}
\end{Schunk}

Since the absorption of chlorophyll is not restricted to the central wavelength, but also affects the neighboring bands, the reflectance values are considerably lowered in the blue and red parts which lead to ``absorption features'' in the spectral signature of the reflectance (shown as gray boxes in Figure \ref{fig:chloro_cr}a). The form and magnitude of these absorption features are correlated to the chlorophyll content of the measured vegetation \citep{Mutanga2004b,Mutanga2004a}. To enhance the form of the absorption features, the spectra can be transformed by constructing a continuum hull around each spectrum. In general, there are two methods for defining such a hull. In the first approach, the convex hull uses the global maximum of the reflectance values as an initial fix point. Then, additional fix points are found to create a convex hull (see red line in Figure \ref{fig:chloro_cr}a). The second approach is called segmented upper hull. Here, the slope of the line to the left and right of the maximum must be positive and negative, respectively (see blue line in Figure \ref{fig:chloro_cr}a). This does not necessarily mean the hull is convex, however. Geologic hyperspectral analyzes often use the convex hull because the distinct absorption features of minerals in the mid-infrared part of the spectrum are easily derived. In vegetation studies, the absorption features of chlorophyll are very close to one  another and the reflectance maximum in the green part is considerably lower than in the near infrared. Consequently, only one absorption feature would be detectable. Therefore, a segmented upper hull (option 'sh') is used in this example to ensure that two small features are identified instead of one large feature. To enhance the chlorophyll absorption features,  the reflectance values are afterward transformed into band depth values (option 'bd'):
\begin{equation}
\mathit{BD}_\mathit{d, \lambda} = 1-\frac{R_\lambda}{\mathit{CV}_\lambda}
\end{equation}
where $R$ is the measured reflectance and $\mathit{CV}$ is the reflectance value of the constructed continuum line at wavelength $\lambda$. 
\begin{Schunk}
\begin{Sinput}
R> spec_bd <- transformSpeclib(spectral_data, method = "sh", out = "bd")
\end{Sinput}
\end{Schunk}

The band depth values in relation to the wavelength of all 45 spectra are plotted in Figure \ref{fig:chloro_cr}b. The chlorophyll absorption features correspond to the first two peaks of the band depth values. The absorption features are now defined as the part of the spectrum between two fix points (band depth values of 0). Since the third absorption feature centered around 980~nm is related to plant water content and biomass rather than chlorophyll \citep{Penuelas1993}, only the absorption features at 460~nm and 670~nm are selected for further analysis.
\begin{Schunk}
\begin{Sinput}
R> featureSpace <- specfeat(spec_bd, c(460, 670))
\end{Sinput}
\end{Schunk}

Several parameters can be calculated from absorption features. These include the wavelength values corresponding to the maximum and the half maximum band depth values. Additionally, the area under the curve is extracted as well as the difference between an idealized Gaussian curve and the observed band depth values. See Table \ref{tab:feat_prop} for a subset of the resulting parameters of the example data set.
\begin{Schunk}
\begin{Sinput}
R> featureSpace <- feature_properties(featureSpace)
\end{Sinput}
\end{Schunk}

\begin{table}[!t]
\resizebox{1\textwidth}{!}{
\begingroup\tiny
\begin{tabular}{rrrrrrrrrrr}
  \toprule
 ID&\multicolumn{2}{c}{Area}&\multicolumn{2}{c}{Width}&\multicolumn{2}{c}{Feature}&\multicolumn{4}{c}{Dist.~to Gauss Curve}\\
 &&&&&\multicolumn{2}{c}{Width}&\multicolumn{2}{c}{$f_{460}$}&\multicolumn{2}{c}{$f_{670}$}\\
 \cmidrule(r){2-3}\cmidrule(r){4-5}\cmidrule(r){6-7}\cmidrule(r){8-9}\cmidrule(r){10-11}
 &$f_{460}$&$f_{670}$&$f_{460}$&$f_{670}$&$f_{460}$&$f_{670}$&left&right&left&right
\\\midrule 
 1 & 23.85 & 131.44 & 518 & 715 & 0.11 & 0.77 & 191.00 & 0.13 & 139.00 & 0.06 \\ 
  2 & 22.13 & 134.01 & 521 & 716 & 0.12 & 0.76 & 194.00 & 0.11 & 142.00 & 0.06 \\ 
  3 & 31.44 & 136.32 & 520 & 718 & 0.11 & 0.78 & 194.00 & 0.13 & 144.00 & 0.07 \\ 
  4 & 17.26 & 132.26 & 519 & 715 & 0.11 & 0.77 & 192.00 & 0.12 & 139.00 & 0.06 \\ 
  5 & 21.75 & 135.03 & 520 & 716 & 0.12 & 0.78 & 193.00 & 0.10 & 142.00 & 0.07 \\ 
  6 & 23.88 & 132.46 & 519 & 717 & 0.11 & 0.76 & 192.00 & 0.12 & 142.00 & 0.06 \\ 
  7 & 21.39 & 136.13 & 519 & 716 & 0.11 & 0.78 & 193.00 & 0.13 & 141.00 & 0.07 \\ 
  8 & 20.75 & 134.76 & 519 & 720 & 0.11 & 0.79 & 193.00 & 0.12 & 147.00 & 0.07 \\ 
  9 & 22.75 & 138.98 & 520 & 717 & 0.12 & 0.80 & 194.00 & 0.12 & 143.00 & 0.07 \\ 
  10 & 22.94 & 130.43 & 520 & 716 & 0.11 & 0.76 & 192.00 & 0.11 & 141.00 & 0.06 \\ 
  11 & 27.89 & 135.50 & 520 & 716 & 0.12 & 0.77 & 193.00 & 0.12 & 142.00 & 0.06 \\ 
  12 & 24.28 & 129.25 & 519 & 718 & 0.11 & 0.76 & 192.00 & 0.12 & 144.00 & 0.06 \\ 
  13 & 26.50 & 135.68 & 520 & 718 & 0.11 & 0.77 & 195.00 & 0.14 & 145.00 & 0.07 \\ 
  14 & 22.13 & 131.74 & 520 & 718 & 0.11 & 0.77 & 193.00 & 0.11 & 144.00 & 0.07 \\ 
  15 & 21.36 & 134.58 & 520 & 717 & 0.12 & 0.77 & 193.00 & 0.12 & 143.00 & 0.06 \\ 
  16 & 37.25 & 123.95 & 514 & 718 & 0.11 & 0.77 & 192.00 & 0.13 & 143.00 & 0.06 \\ 
  17 & 36.99 & 131.96 & 519 & 718 & 0.12 & 0.75 & 193.00 & 0.14 & 146.00 & 0.07 \\ 
  18 & 45.60 & 127.86 & 517 & 719 & 0.11 & 0.75 & 191.00 & 0.15 & 146.00 & 0.06 \\ 
  19 & 42.09 & 130.61 & 518 & 718 & 0.11 & 0.77 & 194.00 & 0.15 & 144.00 & 0.06 \\ 
  20 & 51.52 & 129.11 & 518 & 718 & 0.11 & 0.75 & 190.00 & 0.15 & 145.00 & 0.06 \\ 
  21 & 39.35 & 126.57 & 518 & 718 & 0.11 & 0.73 & 195.00 & 0.13 & 144.00 & 0.06 \\ 
  22 & 47.63 & 130.76 & 517 & 718 & 0.11 & 0.77 & 192.00 & 0.16 & 144.00 & 0.06 \\ 
  23 & 39.94 & 128.55 & 515 & 718 & 0.10 & 0.77 & 194.00 & 0.14 & 143.00 & 0.07 \\ 
  24 & 41.99 & 128.45 & 517 & 718 & 0.11 & 0.76 & 190.00 & 0.15 & 144.00 & 0.06 \\ 
  25 & 48.01 & 128.43 & 518 & 717 & 0.11 & 0.75 & 190.00 & 0.14 & 144.00 & 0.06 \\ 
  26 & 38.35 & 134.08 & 518 & 718 & 0.11 & 0.77 & 193.00 & 0.15 & 145.00 & 0.07 \\ 
  27 & 35.58 & 130.27 & 517 & 719 & 0.10 & 0.75 & 195.00 & 0.14 & 146.00 & 0.06 \\ 
  28 & 45.22 & 131.08 & 517 & 719 & 0.11 & 0.76 & 192.00 & 0.15 & 146.00 & 0.06 \\ 
  29 & 47.61 & 130.07 & 517 & 718 & 0.10 & 0.76 & 194.00 & 0.14 & 144.00 & 0.07 \\ 
  30 & 42.90 & 130.90 & 519 & 719 & 0.12 & 0.75 & 193.00 & 0.15 & 148.00 & 0.07 \\ 
  31 & 50.20 & 128.63 & 520 & 722 & 0.12 & 0.70 & 202.00 & 0.18 & 152.00 & 0.07 \\ 
  32 & 45.42 & 129.62 & 520 & 724 & 0.12 & 0.71 & 202.00 & 0.21 & 155.00 & 0.08 \\ 
  33 & 46.55 & 132.49 & 520 & 721 & 0.12 & 0.72 & 202.00 & 0.21 & 150.00 & 0.07 \\ 
  34 & 46.95 & 133.73 & 521 & 722 & 0.12 & 0.71 & 204.00 & 0.20 & 152.00 & 0.08 \\ 
  35 & 56.06 & 129.62 & 521 & 724 & 0.13 & 0.70 & 203.00 & 0.18 & 156.00 & 0.08 \\ 
  36 & 43.08 & 130.81 & 520 & 722 & 0.12 & 0.70 & 203.00 & 0.21 & 152.00 & 0.07 \\ 
  37 & 36.21 & 135.46 & 521 & 723 & 0.13 & 0.72 & 204.00 & 0.19 & 154.00 & 0.08 \\ 
  38 & 45.62 & 134.72 & 521 & 723 & 0.12 & 0.72 & 203.00 & 0.20 & 154.00 & 0.08 \\ 
  39 & 46.81 & 134.62 & 520 & 722 & 0.12 & 0.74 & 202.00 & 0.22 & 153.00 & 0.08 \\ 
  40 & 46.84 & 134.71 & 520 & 723 & 0.13 & 0.73 & 202.00 & 0.20 & 154.00 & 0.08 \\ 
  41 & 41.39 & 133.68 & 521 & 722 & 0.13 & 0.72 & 204.00 & 0.20 & 153.00 & 0.08 \\ 
  42 & 43.09 & 134.26 & 520 & 723 & 0.12 & 0.73 & 203.00 & 0.21 & 154.00 & 0.08 \\ 
  43 & 50.85 & 130.39 & 520 & 724 & 0.13 & 0.70 & 203.00 & 0.21 & 156.00 & 0.08 \\ 
  44 & 44.85 & 131.95 & 520 & 722 & 0.12 & 0.72 & 202.00 & 0.19 & 153.00 & 0.07 \\ 
  45 & 44.30 & 135.09 & 520 & 722 & 0.13 & 0.73 & 202.00 & 0.21 & 153.00 & 0.07 \\ 
   \bottomrule 
\end{tabular}
\endgroup}
\caption{Selected Feature Properties Extracted from the Band Depth Values. The Area is the Sum of all Band Depth Values within the Respective Feature. The Feature Width is the Difference Between the Wavelength Values at the Upper and Lower FWHM-Values. Distance to Gauss Curve is the Root Mean Square Error (RMSE) of the Part Smaller than (Left) and Greater than (Right) the Maximum. Note that Each Line Represents one Spectral Measurement and the two Chlorophyll Absorption Features are Abbreviated According to Their Central Wavelengths as $f_{460}$ and $f_{670}$.\label{tab:feat_prop}}
\end{table}

In the last part of this example, the chlorophyll contents of the measured samples are estimated using the parameters derived from the absorption feature and the band depth values within the features as predictors. Multivariate statistics and machine learning approaches are frequently used for this purpose, because prediction models based on multiple (and often correlated) variables usually out-perform the univariate approaches. To cope with multivariate and machine learning tasks, \hsdar provides wrapper functions that enable the user to directly use the functionalities of the \pkg{caret} package. This is by far the most comprehensive multivariate package since it includes various approaches with the same syntax and functions. To use the functions of \pkg{caret}, the response variable has to be defined, which must be stored in the SI attached to the Speclib object (``featureSpace'').   
\begin{Schunk}
\begin{Sinput}
R> featureSpace <- setResponse(featureSpace, "chlorophyll")
\end{Sinput}
\end{Schunk}
The spectra are the default selection for predictors. However, additional predictor variables from the attributes of the spectra can be included. In this example, all parameters extracted above are added.
\begin{Schunk}
\begin{Sinput}
R> featureSpace <- setPredictor(featureSpace,
+    names(SI(featureSpace))[4:ncol(SI(featureSpace))])
\end{Sinput}
\end{Schunk}

\begin{figure}[!t]
\centering
\resizebox{0.75\textwidth}{!}{
\includegraphics{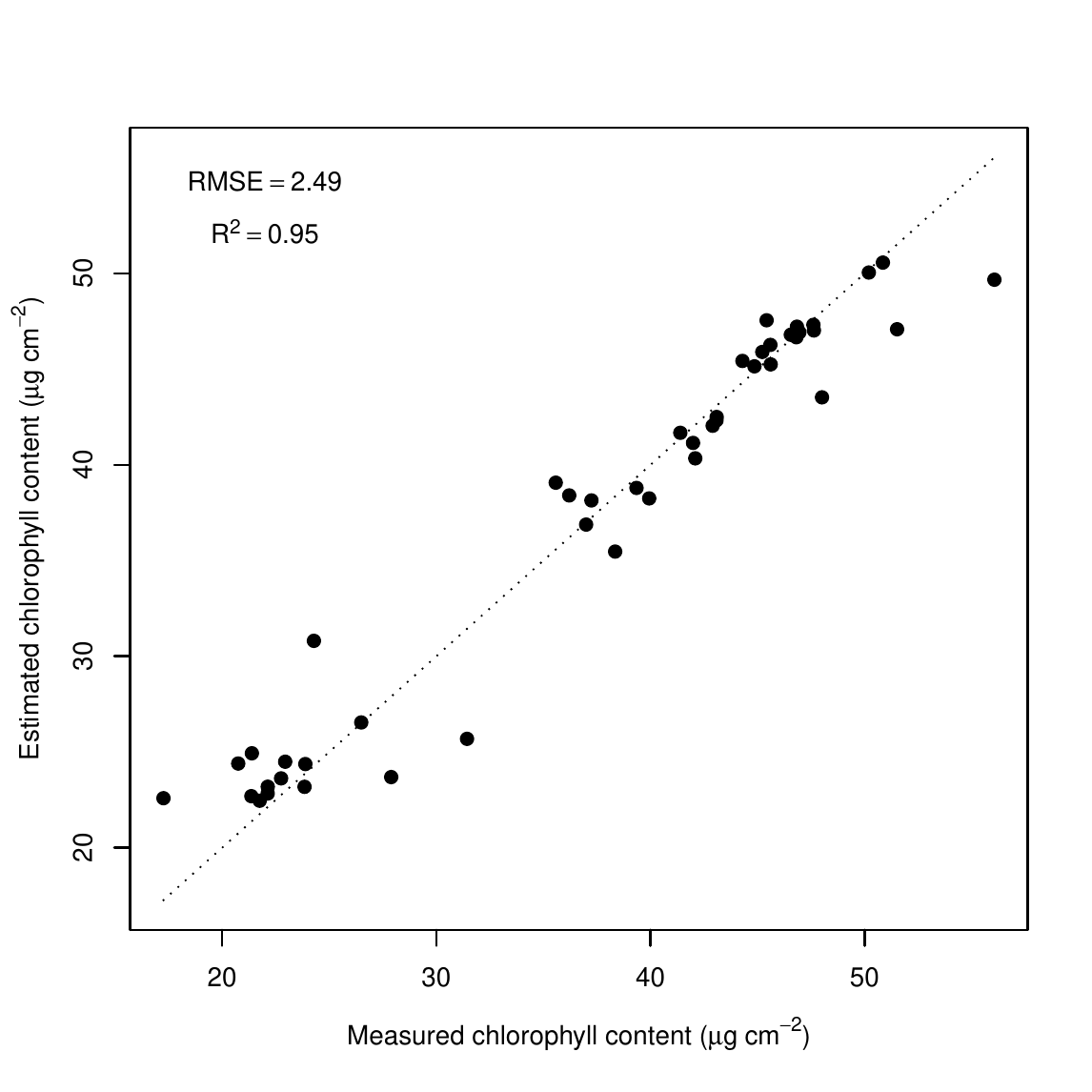}
}
\caption{Estimated vs.~Measured Chlorophyll Content.\label{fig:chl_obe_est}}
\end{figure}

The final model for deriving chlorophyll content is trained by tuning the required parameter for the Random Forest model (Number of randomly selected predictor variables, mtry). 10-fold cross validation is repeated 5 times for model tuning and estimating accuracy. The internal predictions of the final tuning setup are returned providing an independent data set for validation. The accuracy of the predictions performed by the model is evaluated with the \ac{RMSE} and the R$^2$-value. For further information about strategies on model settings and cross validation see \cite{Kuhn2013} and \cite{Kuhn2008}.
\begin{Schunk}
\begin{Sinput}
R> ctrl <- trainControl(method = "repeatedcv", number = 10, repeats = 5, 
+    savePredictions = "final")
R> rfe_trained <- train(featureSpace, trControl = ctrl, method = "rf")
\end{Sinput}
\end{Schunk}
The number of randomly selected predictor variables at each split of the trees is set to mtry = 453. Using the repeated cross validation, the chlorophyll contents estimated by the Random Forest model fit well if compared to the measured ones (\ac{RMSE} = 2.49 mg, $R^2$ = 0.95, Figure \ref{fig:chl_obe_est}). This shows that the proposed method incorporating hyperspectral data is a valid approach for chlorophyll estimation. The resulting model can be used to predict the chlorophyll content of plots where it has not been measured in the field (e.g.,~\citealp{Lehnert2014}).

\subsection{Hyperspectral detection of cancer}
\begin{figure}
\centering
\resizebox{1\textwidth}{!}{
\includegraphics{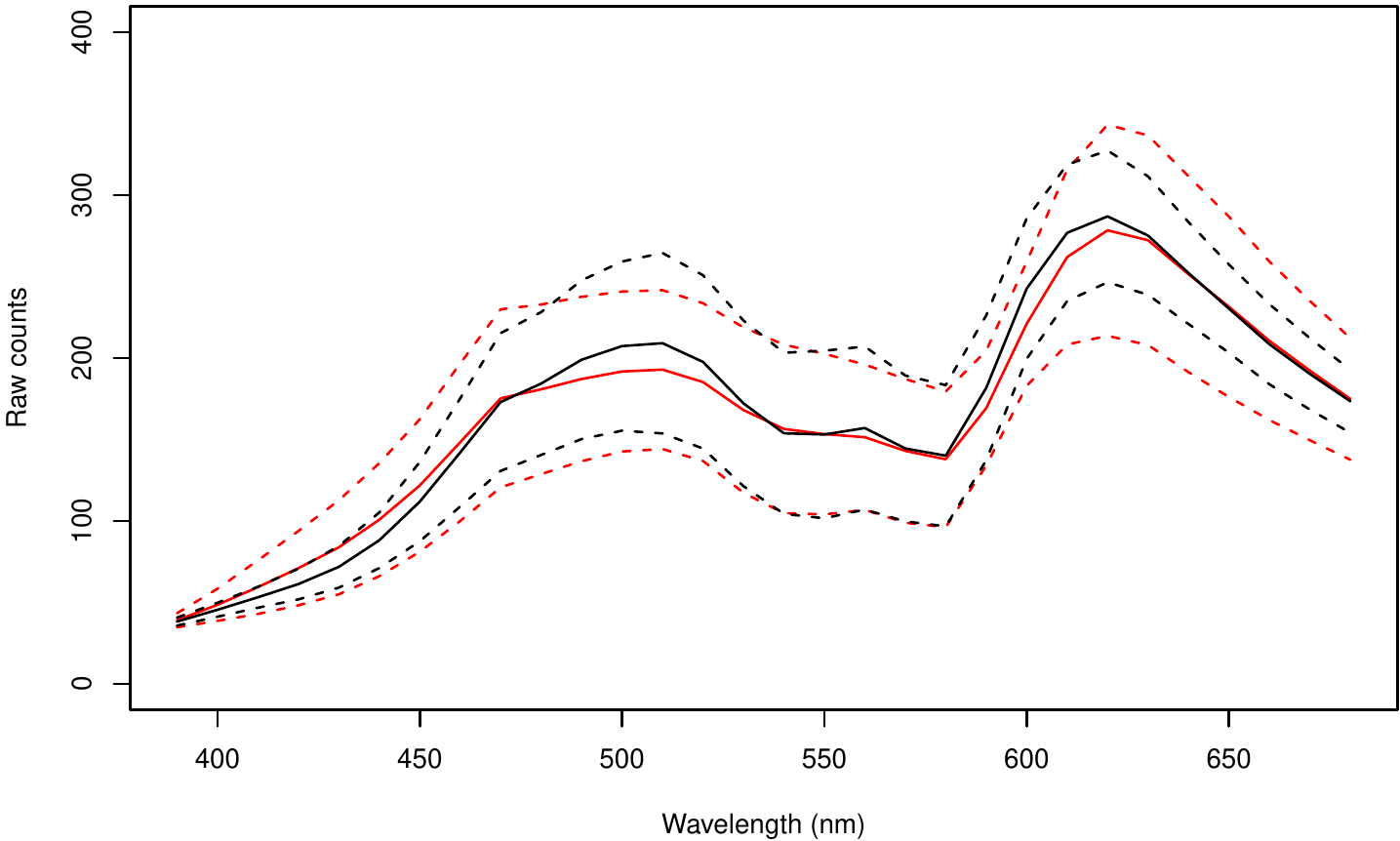}
}
\caption{Spectral Data of the Cancerous (red) and Non-cancerous (black) Parts of the Larynx Showing the Mean (Solid Line) and Standard Deviation (Dashed Lines) of the Count Values Detected by the Monochromatic CCD Camera. \label{fig:cancer_spectra}}
\end{figure}

The second example shows how hyperspectral imaging can be used in non-invasive detection of cancer of the human larynx (head and neck squamous cell carcinoma; hence referred to as ``HNSCC''). This is demonstrated with a data subset acquired at the University of Bonn, Germany that includes hyperspectral images from 25 patients, 10 of which have a histopathological diagnosis of HNSCC. The images were acquired using an endoscope, which was coupled with a monochromatic CCD camera. A special Polychrome V light machine allowed researchers to change the wavelength of the impinging radiation so that several images taken under different illuminations could be combined into hyperspectral cubes (Figure \ref{fig:imgs_exmpl}b). The images were preprocessed and collocated using the methodology proposed by \cite{Regeling2015}. The preprocessing is key because the different bands are acquired with short time lapse as a consequence of the varying light source.  Medical experts' manual classification into cancerous and non-cancerous tissue was used as reference. The following code loads the data into \proglang{R} and plots them to explore the differences between cancerous and non-cancerous tissue (Figure \ref{fig:cancer_spectra}).
\begin{Schunk}
\begin{Sinput}
R> data("cancer_spectra")
R> plot(subset(cancer_spectra, infected == 1), ylim = c(0, 400), 
+    col = "darkred")
R> plot(subset(cancer_spectra, infected == 0), new = FALSE)
\end{Sinput}
\end{Schunk}
Additionally, the response variable (``infected'') is converted to a factor:
\begin{Schunk}
\begin{Sinput}
R> SI(cancer_spectra)$infected <- as.factor(SI(cancer_spectra)$infected)
\end{Sinput}
\end{Schunk}

In contrast to the first example, the spectra of the human larynx are expressed in counts and not reflectance values. Thus, the absolute values highly depend on the light source, the temperature of the sensor, and the illumination geometry. To cope with this limitation, normalized ratio indices are calculated instead of using the absolute count values. Mathematically, these are defined as:
\begin{equation}
\mathit{NRI}_{i,j} = \frac{R_i-R_j}{R_i+R_j}
\end{equation}
Here, $R$ is the reflectance (or in this case the number of counts) at wavelength $i$ or  $j$. These indices are then calculated for all possible combinations of bands through the predefined function ``nri''.
\begin{Schunk}
\begin{Sinput}
R> nri_data <- nri(cancer_spectra, recursive = TRUE)
\end{Sinput}
\end{Schunk}
The $\mathit{NRI}$ values can be directly used as predictors in univariate generalized linear models, for example. Note that a multitude of models must be derived depending on the number of bands in the hyperspectral dataset. Initially, it is worthwhile to resample the spectra to a coarser spectral resolution to reduce the number of models. Alternatively, some functions in \pkg{hsdar} directly support parallel processing using the \pkg{foreach} package. To execute a function on two cores in parallel, simply use the following code depending on the operating system. \newpage
For Linux/Mac OS:
\begin{Schunk}
\begin{Sinput}
R> library("doMC")
R> n_cores <- 2
R> registerDoMC(n_cores)
\end{Sinput}
\end{Schunk}
For Windows:
\begin{Schunk}
\begin{Sinput}
R> library("doMPI")
R> n_cores <- 2
R> cl <- startMPIcluster(count = n_cores)
R> registerDoMPI(cl)
\end{Sinput}
\end{Schunk}
Please note that the dataset in the current example is not large enough to benefit from parallel processing. Therefore, the previous code snippet can be skipped, and we continue by calculating the generalized linear models using the $\mathit{NRI}$ values as predictors for infection:
\begin{Schunk}
\begin{Sinput}
R> glm_models <- glm.nri(infected ~ nri_data, preddata = cancer_spectra, 
+    family = binomial)
\end{Sinput}
\end{Schunk}
It must be noted that the indices are highly correlated, which is a common drawback to using them in a multivariate analysis. In this example, however, each index is used as a predictor in a separate model to eliminate collinearity. 

\begin{figure}[!t]
\centering
\resizebox{0.75\textwidth}{!}{
\includegraphics{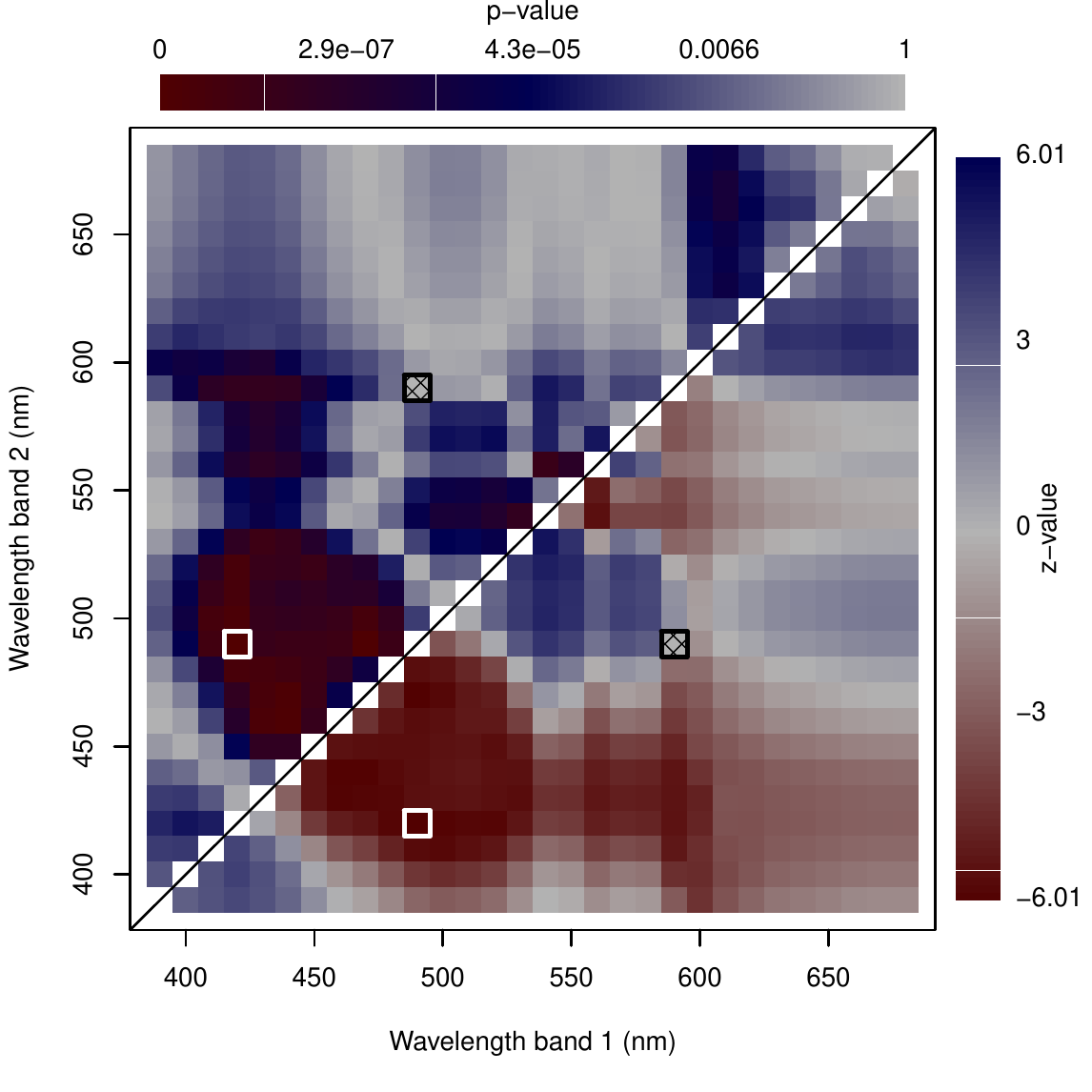}
}
\caption{Relationship between Cancer and Normalized Ratio Indices. The Lower, Right Portion (Triangle) of the Graph Shows the z-values of the Binomial Regression and the Upper Triangle Represents the Corresponding p-values. The White Squares Mark the Positions of the Index (z- and p-values), That Perform Best, While the Black Squares Show the Index With the Worst Performance. Note that Color of p-values is Logarithmically Scaled.\label{fig:cancer_nri}}
\end{figure}

The coefficients, p-values and test statistics of the generalized linear models can now be plotted in 2-d correlograms. In such diagrams, the x-axis and the y-axis represent the two spectral bands used to calculate the index. The color in the diagram symbolizes the coefficient of the model. Thus, the diagrams provide an initial look at band combinations that might be useful for distinguishing between cancerous and non-cancerous parts of the tissue.
\begin{Schunk}
\begin{Sinput}
R> plot(glm_models, coefficient = "z.value", legend = "outer")
R> plot(glm_models, coefficient = "p.value", uppertriang = TRUE, 
+    zlog = TRUE)
\end{Sinput}
\end{Schunk}
The plot is shown in Figure \ref{fig:cancer_nri}.
Almost every index calculated from wavelengths between 400~nm and 450~nm and any other band featured low p-values and, thus, had a significant effect on the distinction between cancerous and non-cancerous tissue (see white rectangle in Figure \ref{fig:cancer_nri}).  Positive z-values were observed for $\mathit{NRI}$ values calculated from longer wavelengths. Negative z-values were obtained for indices calculated from 450~nm to 550~nm for the first band and 400~nm to 480~nm for the second band. The index with the worst performance was calculated from bands 490~nm  and 590~nm (see shaded black rectangle in Figure \ref{fig:cancer_nri}).

This approach, however, precludes multiple $\mathit{NRI}$ values from being used as predictors because they are usually highly correlated, as previously mentioned. Thus, machine learning algorithms classify cancerous cells, as in the first example, because collinearity among predictor variables does not affect their predictive performance. Predictor and response variables have to be defined: As response variable, the column ``infected'' in the SI was used and the $\mathit{NRI}$ values are used as predictors by default. The stage of the cancer is used as an additional predictor variable, because the spectral signal in the early stages of the cancer differs from that in later stages. 
\begin{Schunk}
\begin{Sinput}
R> nri_data <- setResponse(nri_data, "infected")
R> nri_data <- setPredictor(nri_data, "stage")
\end{Sinput}
\end{Schunk}
Unlike the first example, highly correlated predictor variables are excluded before model training by applying a recursive feature elimination, which reduces the computational time. Afterwards, two techniques are used to classify cancerous and non-cancerous tissues: (1) support vector machine \citep{Chang2011,Meyer2014}  and (2) neural network classification \citep{Ripley1996,Venables2002}. 
\begin{Schunk}
\begin{Sinput}
R> sel_feat <- rfe(nri_data, cutoff = 0.9)
R> ctrl <- trainControl(method = "repeatedcv", number = 10, repeats = 5,
+    savePredictions = "final")                     
R> rfe_trained_svm  <- train(sel_feat, trControl = ctrl, 
+    importance = TRUE, method = "svmRadial")
R> rfe_trained_nnet <- train(sel_feat, trControl = ctrl,
+    importance = TRUE, method = "nnet")
\end{Sinput}
\end{Schunk}
\begin{table}

\centering
%
%

\begin{minipage}[t]{0.4\textwidth}
\textbf{a}\\
\resizebox{1\textwidth}{!}{
\begin{tabular}{rrr}
  \toprule 
 & Infected & Not Infected  \\\midrule 
 Infected & 68.40 & 3.40 \\ 
  Not Infected & 6.60 & 71.60 \\ 
   \bottomrule 
\end{tabular}}
\end{minipage}\hspace{0.05\textwidth}
\begin{minipage}[t]{0.4\textwidth}
\textbf{b}\\
\resizebox{1\textwidth}{!}{
\begin{tabular}{rrr}
  \toprule 
 & Infected & Not Infected  \\\midrule 
 Infected & 65.60 & 5.60 \\ 
  Not Infected & 9.40 & 69.40 \\ 
   \bottomrule 
\end{tabular}}
\end{minipage}
\caption{Error Matrix of the Obtained Classification Results for the Support Vector Machine (a) and the Neural Network (b) Models. The   Rows and Columns are the Mean Values of Observations and Estimations within the 5 Repeats of the 10-fold Cross Validation, respectively.\label{tab:cancer_vali}}

\end{table}
Table \ref{tab:cancer_vali} shows the validation result of the final models for both methods. Support vector machine performed slightly better and yielded an overall accuracy of 93.33\% as compared to 90\% for the neural network classification. This shows that hyperspectral imaging and machine learning approaches may yield positive results for detecting cancer in human tissue. The data used in this case study have several drawbacks mainly due to the acquisition with a variable light source instead of a hyperspectral camera in combination with a constant light source. This causes the count values to be dependent on movements of the patient and the illumination geometry by the light source. However, the analysis based on normalized ratio indices yielded robust results clearly highlighting its large potential. Since hyperspectral imaging is a non-invasive measurement technology, the examination is relatively comfortable for the patient. However, it has to be noted that the detection of cancer with hyperspectral imaging may only facilitate the diagnose of a medical expert. At the moment, there is no possibility to automatically diagnose cancer in the human larynx without the  knowledge of a trained medical expert \citep{Regeling2016}.

\section{Conclusions}
The two case studies provide an initial impression of what hyperspectral remote sensing can be used for and how a typical approach may look. Both examples show how the \hsdar package can be used as a powerful tool within \proglang{R} for remote sensing and spatial applications. Based on the widely used raster package, \hsdar introduces new functionalities for processing hyperspectral data and gives users control over the results of univariate and multivariate modeling approaches, including machine learning techniques. Although \hsdar is dedicated to spectral data featuring many bands, it is applicable to any multispectral satellite data including Landsat 8 (8 bands in the visible and near infrared part of the electromagnetic radiation) or MODIS (19 bands) \citep{Lehnert2015}. For example, \hsdar can perform linear spectral unmixing or calculate spectral indices such as the NDVI. \hsdar differentiates itself from the other hyperspectral package available for \proglang{R} (\pkg{hyperSpec}, \citealp{Beleites2016}) by focusing on environmental instead of laboratory analysis. Data can easily be transferred between both packages since \hsdar provides functions to convert to and from objects in \pkg{hyperSpec}. Both packages extend \proglang{R} by functions for all state of the art methods in hyperspectral imaging which have been available only in commercial software tools so far.

\section*{Acknowledgments}
Initial development of the \hsdar package was financially supported by the German Federal Ministry of Education and Research (BMBF) within the Pasture Degradation Monitoring System (PaDeMoS) project (03G0808C). Data for the first case study was taken in the framework of  the LOEWE excellence cluster FACE$_2$FACE funded by the Hessian State Ministry of Higher Education, Research and the Arts. The second case study was based on data from the project "Early Detection of Laryngeal Cancer by Hyperspectral Imaging" (German Cancer Aid, project number 109825 and 110275).

\bibliography{jss2837.bib}
\end{document}